\documentclass[12pt]{article}

\usepackage{epsfig}
\usepackage{graphics}
\input epsf

\title{Beauty photoproduction at HERA: $k_T$-factorization \\ versus experimental data}

\author{A.V.~Lipatov, N.P.~Zotov}

\begin{document}

\maketitle

\begin{center}

{\it D.V.~Skobeltsyn Institute of Nuclear Physics,\\ 
M.V. Lomonosov Moscow State University,
\\119992 Moscow, Russia\/}\\[3mm]

\end{center}

\vspace{1cm}

\begin{center}

{\bf Abstract }

\end{center}

We present calculations of the beauty photoproduction at HERA collider 
in the framework of the $k_T$-factorization approach. Both direct and 
resolved photon contributions are taken into account. 
The unintegrated gluon densities in a proton and in a 
photon are obtained from the full CCFM, from unified BFKL-DGLAP evolution 
equations as well as from the Kimber-Martin-Ryskin prescription.
We investigate different production rates (both inclusive and associated 
with hadronic jets) and compare our theoretical predictions with the recent 
experimental data taken by the H1 and ZEUS collaborations.
Special attention is put on the $x_\gamma^{\rm obs}$ variable which 
is sensitive to the relative contributions to the beauty production cross section. 

\vspace{1cm}

\section{Introduction} \indent 

The beauty production at high energies is a subject
of intensive study from both theoretical and experimental
points of view. First measurements~[1] of the $b$-quark cross
sections at HERA were significantly higher than QCD predictions 
calculated at next-to-leading order (NLO). 
Similar observations were made in hadron-hadron
collisions at Tevatron~[2] and also in photon-photon interactions
at LEP2~[3]. In last case, the theoretical NLO QCD predictions are
more than three standard deviations below the experimental data.
At Tevatron, recent analisys indicates that the overall description 
of the data can be improved~[4] by adopting the non-perturbative 
fragmentation function of the $b$-quark into the $B$-meson: an appropriate
treatment of the $b$-quark fragmentation properties considerably
reduces the disagreement between measured beauty cross section
and the corresponding NLO QCD calculations. 
Recently H1 and ZEUS collaborations have reported important data~[5--7] on the 
beauty photoproduction (both inclusive and associated with hadronic jets) in 
electron-proton collisions at HERA which refer
to small values of the Bjorken scaling variable $x$.
These data are in a reasonable agreement 
with NLO QCD predictions or somewhat higher. Some disagreement is observed~[7] 
mainly at small decay muon and/or associated jet transverse momenta.
But the large excess of the first measurements over NLO QCD, reported~[1]
by the H1 collaboration, is not confirmed.
In the present paper to analyze the H1 and ZEUS 
data we will apply the so-called $k_T$-factorization~[8, 9] (or semi-hard~[10, 11]) 
approach of QCD since the beauty production at HERA is 
dominated by the photon-gluon or gluon-gluon fusion 
(direct and resolved photon contributions, respectively) and therefore 
sensitive to the gluon densities in a proton and in a photon at
small values of $x$.

The $k_T$-factorization approach is based on the Balitsky-Fadin-Kuraev-Lipatov 
(BFKL)~[12] or Ciafaloni-Catani-Fiorani-Marchesini (CCFM)~[13] gluon evolution
which are valid at small $x$ since here large logarithmic terms proportional to $\ln 1/x$
are summed up to all orders of perturbation theory (in the leading logarithmic approximation). 
It is in contrast with the popular 
Dokshitzer-Gribov-Lipatov-Altarelli-Parizi (DGLAP)~[14]
strategy where only large logarithmic terms proportional 
to $\ln \mu^2$ are taken into account. The basic dynamical quantity of the 
$k_T$-factorization approach is the so-called unintegrated 
(${\mathbf k}_T$-dependent) gluon distribution 
${\cal A}(x,{\mathbf k}_T^2,\mu^2)$ which determines the probability to find a gluon carrying 
the longitudinal momentum fraction $x$ and the transverse momentum 
${\mathbf k}_T$ at the probing scale $\mu^2$. 
The unintegrated gluon distribution can be obtained from the
analytical or numerical solution of the BFKL or CCFM evolution equations.
Similar to DGLAP, to calculate the cross sections of any physical process the unintegrated 
gluon density ${\cal A}(x,{\mathbf k}_T^2,\mu^2)$ has to be convoluted~[8--11] with the relevant 
partonic cross section $\hat \sigma$. But as the virtualities of the propagating
gluons are no longer ordered, the partonic cross section has
to be taken off mass shell (${\mathbf k}_T$-dependent). It is in clear contrast 
with the DGLAP scheme (so-called collinear factorization). Since gluons in initial 
state are not on-shell and are characterized by virtual masses 
(proportional to their transverse momentum), it also assumes a modification 
of their polarization density matrix~[10, 11]. In particular, the polarization vector of a gluon is no
longer purely transversal, but acquires an admixture of longitudinal 
and time-like components. Other important properties of the $k_T$-factorization 
formalism are the additional contribution to the cross sections due to
the integration over the ${\mathbf k}_T^2$ region above $\mu^2$
and the broadening of the transverse momentum distributions due to extra 
transverse momentum of the colliding partons.

Some applications of the $k_T$-factorization approach supplemented with 
the BFKL and CCFM evolution to the $b$-quark production at high energies 
were discussed in~[15--26]. It was shown~[17--21] that the beauty cross section at 
Tevatron can be consistently described in the framework of this 
approach. However, a substantial discrepancy between theory
and experiment is still found~[22--25] for the $b$-quark production in 
$\gamma \gamma$ collisions at LEP2, not being cured by
the $k_T$-factorization\footnote{Some discussions of this problem 
may be found in~[22, 24].}. At HERA, the inclusive beauty photoproduction
has been investigated~[18, 22, 26] and comparisons with the 
first H1 measurements~[1] have been done. It was concluded that the $k_T$-factorization
approach provide a reasonable description of the data within the
large theoretical and experimental uncertainties.
In~[18, 26] the Monte-Carlo 
generator \textsc{Cascade}~[27] has been used 
to predict the cross section of the $b$-quark and dijet associated 
photoproduction. However, all calculations~[18, 22, 26] deal with the
total cross sections only. A number of important differential 
cross sections (such as transverse momentum and pseudo-rapidity 
distributions) has not been considered and comparisons with the 
recent H1 and ZEUS measurements~[5--7] have not been made.

In the present paper we will study the beauty production at HERA 
in more detail. We investigate a number of different photoproduction 
rates (in particular, the transverse momentum and pseudo-rapidity 
distributions of muons which originate from the semi-leptonic decays of $b$-quarks)
and make comparisons with the recent H1 and ZEUS data~[5--7].
Both direct ($\gamma g \to b\bar b$) and resolved photon contributions 
($gg \to b\bar b$) will be taken into account\footnote{The $b$-quark 
excitation processes $bg \to bg$ are automatically included in the $k_T$-factorization 
approach, as it was demonstrated in~[28--30].}. Our analysis covers also 
both inclusive and dijet associated  
$b$-quark production. In last case special attention will be 
put on the $x_\gamma^{\rm obs}$ variable since this quantity is 
sensitive to the relative contributions to the cross section from different 
production mechanisms. One of the purposes of this paper is to investigate the 
specific $k_T$-factorization effects in the $b$-quark production at HERA.
In the numerical analysis we test the unintegrated gluon 
distributions which are obtained from the full CCFM, unified BFKL-DGLAP~[31] evolution 
equations and from the conventional parton densities (using the 
Kimber-Martin-Ryskin prescription~[32]). We would like to note that this study is the  
continuation of our previous investigations~[24, 30] where we have discussed, in particular, 
the charm production at LEP2~[24] and HERA~[30].

The outline of  our paper is following. In Section~2 we 
recall the basic formulas of the $k_T$-factorization approach with a brief 
review of calculation steps. In Section~3 we present the numerical results
of our calculations and a dicussion. Finally, in Section~4, we give
some conclusions.

\section{Basic formulas} 
\subsection{Kinematics} \indent 

We start from the gluon-gluon fusion subprocess. 
Let $p_e$ and $p_p$ be the four-momenta of the initial electron and 
proton, $k_1$ and $k_2$ the four-momenta of the incoming off-shell gluons, 
and $p_b$ and $p_{\bar b}$ the four-momenta of the produced
beauty quarks. In our analysis below we will use the Sudakov 
decomposition, which has the following form:
$$
  p_b = \alpha_1 p_e + \beta_1 p_p + p_{b\, T},\quad p_{\bar b} = \alpha_2 p_e + \beta_2 p_p + p_{\bar b\, T},\atop
  k_1 = x_1 p_e + k_{1T},\quad k_2 = x_2 p_p + k_{2T}, \eqno(1)
$$

\noindent 
where $k_{1T}$, $k_{2T}$, $p_{b\, T}$ and $p_{\bar b\, T}$ are the
transverse four-momenta of the corresponding particles.
It is important that ${\mathbf k}_{1T}^2 = - k_{1T}^2 \neq 0$ and
${\mathbf k}_{2T}^2 = - k_{2T}^2 \neq 0$. If we make replacement $k_1 \to p_e$
and set $x_1 = 1$ and $k_{1T} = 0$, then we easily obtain more
simpler formulas corresponding to photon-gluon fusion subprocess.
In the $ep$ center-of-mass frame we can write
$$
  p_e = {\sqrt s}/2 (1,0,0,1),\quad p_p = {\sqrt s}/2 (1,0,0,-1), \eqno(2)
$$

\noindent
where $s = (p_e + p_p)^2$ is the total energy of the process under consideration
and we neglect the masses of the incoming particles. The Sudakov variables
are expressed as follows:
$$
  \displaystyle \alpha_1={m_{b\, T}\over {\sqrt s}}\exp(y_b),\quad \alpha_2={m_{\bar b\, T}\over {\sqrt s}}\exp(y_{\bar b}),\atop
  \displaystyle \beta_1={m_{b\, T}\over {\sqrt s}}\exp(-y_b),\quad \beta_2={m_{\bar b\, T}\over {\sqrt s}}\exp(-y_{\bar b}), \eqno(3)
$$

\noindent
where $m_{b\, T}$ and $m_{\bar b\, T}$ are the transverse masses 
of the produced quarks, and $y_b$ and $y_{\bar b}$ 
are their rapidities (in the $ep$ center-of-mass frame). 
From the conservation laws we can easily obtain the following conditions:
$$
  x_1 = \alpha_1 + \alpha_2,\quad x_2 = \beta_1 + \beta_2,\quad {\mathbf k}_{1T} + {\mathbf k}_{2T} = {\mathbf p}_{b\, T} + {\mathbf p}_{\bar b \, T}. \eqno(4)
$$

The variable $x_\gamma^{\rm obs}$ is often used
in the analysis of the data which contain the dijet samples. 
This variable, which is the fraction of the photon momentum contributing 
to the production of two hadronic jets with transverse energies $E_T^{{\rm jet}_1}$
and $E_T^{{\rm jet}_2}$, experimentally is defined~[6, 7] as
$$
  x_\gamma^{\rm obs} = { E_T^{{\rm jet}_1} e^{-\eta^{{\rm jet}_1}} + E_T^{{\rm jet}_2} e^{-\eta^{{\rm jet}_2}} \over 2 y E_e}, \eqno (5)
$$

\noindent
where $y E_e$ is the initial photon energy and $\eta^{{\rm jet}_i}$ are
the pseudo-rapidities of these jets. The pseudo-rapidities $\eta^{{\rm jet}_i}$ 
are defined as $\eta^{{\rm jet}_i} = - \ln \tan (\theta^{{\rm jet}_i}/2)$, where 
$\theta^{{\rm jet}_i}$ are the polar angles of the jets with respect to the proton beam.
Note that the selection of $x_\gamma^{\rm obs} > 0.75$ and $x_\gamma^{\rm obs} < 0.75$ 
yields samples enriched in direct and resolved photon processes, respectively.

\subsection{Cross section for beauty photoproduction} \indent 

The main formulas for the total and differential beauty production cross sections 
were obtained in our previous papers~[19, 24]. Here we recall some of them.
In general case, the cross section $\sigma$ 
according to $k_T$-factorization theorem can be written as a convolution
$$
  \sigma = \sigma_0 \int {dz\over z} \, d{\mathbf k}_{T}^2 \, C(x/z,{\mathbf k}_{T}^2,\mu^2) {\cal A}(x,{\mathbf k}_{T}^2,\mu^2), \eqno(6)
$$

\noindent
where $C(x,{\mathbf k}_{T}^2,\mu^2)$ is the coefficient function
corresponding to relevant partonic subprocess under consideration. So, the direct photon 
contribution to the differential cross section of $\gamma p \to b\bar b + X$ process is given by
$$
  { d\sigma^{\rm (dir)} (\gamma p \to b\bar b + X) \over dy_b\, d{\mathbf p}_{b\, T}^2 } =
  \int {|\bar {\cal M}|^2(\gamma g^* \to b\bar b)\over 16\pi (x_2 s)^2 (1 - \alpha_1)} {\cal A}(x_2,{\mathbf k}_{2T}^2,\mu^2) d{\mathbf k}_{2T}^2 {d\phi_2 \over 2\pi} {d\phi_b \over 2\pi}, \eqno (7)
$$

\noindent
where $|\bar {\cal M}|^2(\gamma g^* \to b\bar b)$ is the squared off-shell matrix element which 
depends on the transverse momentum ${\mathbf k}_{2T}^2$, $\phi_2$ and $\phi_b$ are the 
azimuthal angles of the initial virtual gluon and the produced quark, respectively.
The formula for the resolved photon contribution can be obtained by the similar way. But
one should keep in mind that convolution in (6) should be made also with the
unintegrated gluon distribution ${\cal A}_\gamma(x,{\mathbf k}_{T}^2,\mu^2)$ in
a photon. The final expression for the differential cross section has the form
$$
  \displaystyle { d\sigma^{\rm (res)} (\gamma p \to b\bar b + X) \over dy_b\, d{\mathbf p}_{b\, T}^2 } = \int {|\bar {\cal M}|^2(g^* g^* \to b\bar b)\over 16\pi (x_1 x_2 s)^2} \times \atop 
  \displaystyle \times {\cal A}_\gamma(x_1,{\mathbf k}_{1T}^2,\mu^2) {\cal A}(x_2,{\mathbf k}_{2T}^2,\mu^2) d{\mathbf k}_{1T}^2 d{\mathbf k}_{2T}^2 dy_{\bar b} {d\phi_1\over 2\pi} {d\phi_2\over 2\pi} {d\phi_b\over 2\pi}, \eqno(8)
$$

\noindent
where $\phi_1$ is the azimuthal angle of the initial virtual gluon having fraction
$x_1$ of a initial photon longitudinal momentum. It is important that squared off-shell matrix 
element $|\bar {\cal M}|^2(g^* g^* \to b\bar b)$ depends on the both transverse momenta 
${\mathbf k}_{1 T}^2$ and ${\mathbf k}_{2 T}^2$. 
The analytic expressions for the $|\bar {\cal M}|^2 (\gamma g^* \to b\bar b)$
and $|\bar {\cal M}|^2 (g^* g^* \to b\bar b)$ have been evaluated in our 
previous papers~[19, 24]. Note that if we 
average (7) and (8) over ${\mathbf k}_{1 T}$ and ${\mathbf k}_{2 T}$ and 
take the limit ${\mathbf k}_{1 T}^2 \to 0$ and ${\mathbf k}_{2 T}^2 \to 0$,
then we obtain well-known formulas corresponding to the leading-order (LO) QCD calculations.

The recent experimental data~[5--7] taken by the H1 and ZEUS collaborations 
refer to the $b$-quark photoproduction in $ep$ collisions, where electron is scattered
at small angle and the mediating photon is almost real ($Q^2 \sim 0$).
Therefore $\gamma p$ cross sections (7) and (8) needs to be weighted with the photon flux 
in the electron:
$$
  d\sigma(ep \to b\bar b + X) = \int f_{\gamma/e}(y)dy\, d\sigma(\gamma p \to b\bar b + X), \eqno (9)
$$

\noindent
where $y$ is a fraction of the initial electron energy taken by the photon in the 
laboratory frame, and we use the Weizacker-Williams approximation for the 
bremsstrahlung photon distribution from an electron:
$$
  f_{\gamma/e}(y) = {\alpha_{em} \over 2\pi}\left({1 + (1 - y)^2\over y}\ln{Q^2_{\rm max}\over Q^2_{\rm min}} + 
  2m_e^2 y\left({1\over Q^2_{\rm max}} - {1\over Q^2_{\rm min}} \right)\right). \eqno (10)
$$

\noindent
Here $\alpha_{em}$ is Sommerfeld's fine structure constant, $m_e$ is the electron 
mass, $Q^2_{\rm min} = m_e^2y^2/(1 - y)^2$ and $Q^2_{\rm max} = 1\,{\rm GeV}^2$, 
which is a typical value for the recent photoproduction measurements at HERA.

The multidimensional integration in (7), (8) and (9) has been performed
by means of the Monte Carlo technique, using the routine 
\textsc{Vegas}~[33]. The full C$++$ code is available from the authors on 
request\footnote{lipatov@theory.sinp.msu.ru}. This code is
practically identical to that used in~[30], with exception
that now we apply it to calculate beauty production instead charm.

\section{Numerical results} \indent 

We now are in a position to present our numerical results. 
First we describe our theoretical input and the kinematical conditions. 

\subsection{Theoretical uncertainties} \indent 
 
There are several parameters which determined the normalization factor of 
the cross sections (7) and (8): the beauty 
mass $m_b$, the factorization and 
normalisation scales $\mu_F$ and $\mu_R$ and the unintegrated gluon distributions 
in a proton ${\cal A}(x,{\mathbf k}_T^2,\mu^2)$ and in a photon 
${\cal A}_\gamma(x,{\mathbf k}_T^2,\mu^2)$. 

Concerning the unintegrated gluon densities in a proton, in the numerical calculations 
we used five different sets of them, namely the J2003 (set~1 --- 3)~[21], KMS~[31]
and KMR~[32]. All these distributions are widely discussed in the literature 
(see, for example, review~[34, 35] for more information). Here we only shortly 
discuss their characteristic properties.
First, three sets of the J2003 gluon density have been obtained~[21] 
from the numerical solution of the full CCFM equation. The input parameters were fitted to 
describe the proton structure function $F_2(x,Q^2)$.
Note that the J2003~set~1 and J2003 set~3 densities contain only
singular terms in the CCFM splitting function $P_{gg}(z)$. 
The J2003~set~2 gluon density takes into account the additional non-singlular 
terms\footnote{See Ref.~[21] for more details.}. These distributions have been 
applied in the analysis of the forward jet production at HERA and charm and bottom 
production at Tevatron~[21] (in the framework of Monte-Carlo generator \textsc{Cascade}) 
and have been used also in our calculations~[30].

Another set (the KMS)~[31] was obtained from a unified 
BFKL-DGLAP description of $F_2(x, Q^2)$ data and includes the so-called 
consistency constraint~[36]. The consistency constraint introduces a 
large correction to the LO BFKL equation. It was argued~[37] that 
about 70\% of the full NLO corrections to the BFKL exponent 
$\Delta$ are effectively included in this constraint. 
The KMS gluon density is successful in description of the 
beauty production at Tevatron~[17, 19] and $J/\psi$ meson photo- and leptoproduction
at HERA~[38, 39].

The last, fifth unintegrated gluon distribution ${\cal A}(x,{\mathbf k}_T^2,\mu^2)$
used here (the so-called KMR distribution) 
is the one which was originally proposed in~[32]. The KMR approach is the formalism 
to construct unintegrated gluon distribution from the known conventional parton
(quark and gluon) densities. It accounts for the angular-ordering (which comes from the coherence 
effects in gluon emission) as well as the main part of the collinear higher-order QCD 
corrections. The key observation here is that the $\mu$ dependence of the unintegrated 
parton distribution enters at the last step of the evolution, and therefore single scale 
evolution equations (DGLAP or unified BFKL-DGLAP) can be used up to this step. 
Also it was shown~[32] that the unintegrated distributions obtained via unified BFKL-DGLAP 
evolution are rather similar to those based on the pure DGLAP equations.
It is because the imposition of the angular ordering constraint is more 
important~[32] than including the BFKL effects. Based on this point, 
in our further calculations we use much more simpler DGLAP equation up to 
the last evolution step\footnote{We have used the standard GRV (LO) parametrizations~[40]
of the collinear quark and gluon densities.}. Note that the KMR parton densities 
in a proton were used, in particular, to describe the prompt photon photo- and hadroproduction 
at HERA~[41] and Tevatron~[42, 43].

In the case of a real photon, we have tested two different sets of the
unintegrated gluon densities ${\cal A}_\gamma(x,{\mathbf k}_T^2,\mu^2)$.
First of them was obtained~[23] from the 
numerical solution of the full CCFM equation (which has been also
formulated for the photon). Here we will use this gluon density together 
with the three sets of the J2003 distribution when calculating the resolved photon 
contribution (8). Also in order to obtain the unintegrated 
gluon density in a photon we will apply the KMR 
method to the standard GRV parton distributions~[40].
In the numerical calculations we will use it together with the KMR distributions
in a proton. Note that both gluon densities ${\cal A}_{\gamma}(x,{\mathbf k}_T^2,\mu^2)$
discussed here have been already applied in the analysis of the heavy 
(charm and beauty) quark~[22--25] and $J/\psi$ meson~[24, 25] production in $\gamma \gamma$ 
collisions at LEP2.

We would like to point out that at present there is not the unintegrated gluon
distribution corresponding to the unified BFKL-DGLAP evolution in a photon. 
Therefore we will not take into account the 
resolved photon contribution (8) in the case of KMS gluon. 

Also the significant theoretical uncertainties in our results connect with the
choice of the factorization and renormalization scales. First of them
is related to the evolution of the gluon distributions, the other is 
responsible for the strong coupling constant $\alpha_s(\mu^2_R)$.
The optimal values of these scales are such that the contribution of higher
orders in the perturbative expansion is minimal.
As it often done~[11, 15--26, 44] for beauty production, we choose the renormalization and 
factorization scales to be equal: $\mu_R = \mu_F = \mu = \sqrt{m_b^2 + \langle {\mathbf p}_{T}^2 \rangle}$, 
where $\langle {\mathbf p}_{T}^2 \rangle$ is set to the average ${\mathbf p}_{T}^2$ 
of the beauty quark and antiquark\footnote{We use special choice $\mu^2 = {\mathbf k}_T^2$ 
in the case of KMS gluon, as it was originally proposed in~[31].}. 
Note that in the present paper we concentrate mostly on the non-collinear gluon 
evolution in the proton and do not study the scale dependence of our 
results. To completeness, we take the $b$-quark mass 
$m_b = 4.75$~GeV and use LO formula for the coupling constant $\alpha_s(\mu^2)$ 
with $n_f = 4$ active quark flavours at $\Lambda_{\rm QCD} = 200$~MeV, such 
that $\alpha_s(M_Z^2) = 0.1232$.

\subsection{Inclusive beauty photoproduction} \indent 

The recent experimental data~[5--7] for the inclusive beauty photoproduction at HERA
comes from both the H1 and ZEUS collaborations. The $b$-quark total cross section
for $p_T > p_T^{\rm min}$ as well as the beauty transverse momentum
distribution have been determined. The ZEUS data~[5, 6] refer to the
kinematical region\footnote{Here and in the following 
all kinematic quantities are given in the laboratory frame where positive OZ axis 
direction is given by the proton beam.} defined by $|\eta^b| < 2$ and $Q^2 < 1$~GeV$^2$,
where $\eta^b$ is the beauty pseudo-rapidity.
The fraction $y$ of the electron energy transferred to the photon
is restricted to the range $0.2 < y < 0.8$.

In Figs.~1 and~2 we show our predictions in comparison to the ZEUS data~[5, 6].
The solid, dashed, dash-dotted, dotted and short dash-dotted curves
correspond to the results obtained with the J2003~set~1 --- 3, KMR and 
KMS unintegrated gluon densities, respectively. One can see that overall agreement
between our predictions and experimental data is a very good. All three 
sets of the J2003 distribution as well as the KMS gluon density give results which are rather close to
each other (except large $p_T^b$ region where the KMS density predicts 
more hard behaviour). We find also a some enhancement of the estimated cross sections 
as compared with the collinear NLO QCD calculations which lie 
somewhat below the measurements but still agree with the data within the scale uncertainties. 
This enhancement comes, in particular, from the non-zero
transverse momentum of the incoming off-shell gluons. Note that
the KMR gluon distribution gives results which lie below the ZEUS data and which 
are very similar to the NLO QCD predictions.
This observation coincides with the ones~[41].
Such underestimation can be explained by the fact that leading logarithmic terms 
proprtional to $\ln 1/x$ are not included into the KMR formalism.

Also the total inclusive beauty cross section $\sigma(ep \to e b\bar b + X)$ 
has been measured~[1] by the H1 collaboration and it was 
found to be equal to $14.8 \pm 1.3$~(stat.)~$^{+3.3}_{-2.8}$~(sys.)~nb for $Q^2 < 1$~GeV$^2$. 
The collinear NLO QCD calculations
predict a cross section which is about a factor of~4 below the H1 measurements~[1].
The results of our calculations supplemented with the different unintegrated gluon densities 
are collected in Table~1. Also the predictions of the Monte-Carlo generator \textsc{Cascade}~[26]
are shown for comparison. One can see that earlier H1 data~[1] exceed our theoretical 
estimations by a factor about~2. However, recent analysis which has been 
performed in~[6, 7] does not confirm the results of the first measurements~[1]. So, the
cross section for muon coming from $b$ decays in dijet photoproduction events was
found~[7] to be significantly lower than one reported in~[1]. 
Therefore we can expect that the inclusive $b$-quark cross section 
(which can be obtained after extrapolation of dijet and muon cross section
to the full phase space) will be reduced and agreement
with our predictions will be significantly improved. 

In general, we can conclude that the cross sections of inclusive 
beauty photoproduction calculated in the $k_T$-factorization
formalism (supplemented with the CCFM or unified BFKL-DGLAP evolution)
are larger by $30 - 40$\% than ones calculated at 
NLO level of collinear QCD. Our results for the total and differential 
cross sections are in a better agreement with the H1 and ZEUS data than 
the NLO QCD predictions.
We find also that the individual contributions from the photon-gluon and 
gluon-gluon fusion to the inclusive $b$-quark cross section in the 
$k_T$-factorization approach is about 85 and 15\%, respectively. 
This is in agreement with the results presented in~[22] where the KMR and GBW 
unintegrated gluon densities has been used.

\begin{table}
\begin{center}
\begin{tabular}{|l|c|}
\hline
  Source & $\sigma(ep \to e' b\bar b + X)$~[nb] \\
\hline
  H1 measurement~[1] & $14.8~\pm~1.3$~(stat.)~$^{+3.3}_{-2.8}$~(sys.) \\
  \textsc{Cascade}~[26] & $5.2^{+1.1}_{-0.9}$ \\
  J2003 set 1 & 6.78 \\
  J2003 set 2 & 6.62 \\
  J2003 set 3 & 7.16 \\
  KMR & 3.91 \\
  KMS & 7.57 \\
\hline
\end{tabular}
\end{center}
\caption{The total cross section of the inclusive beauty photoproduction in electron-proton 
collisions at $Q^2 < 1$~GeV$^2$.}
\end{table}

\subsection{Dijet associated beauty photoproduction} \indent 

Now we demonstrate how the $k_T$-factorization approach can be 
used to calculate the semi-inclusive beauty photoproduction rates.
The basic photon-gluon or gluon-gluon fusion subprocesses 
give rise to two high-energy $b$-quarks, which can further evolve into hadron jets.
In our calculations the produced quarks (with their known kinematical 
parameters) were taken to play the role of the final jets.
These two quarks are accompanied by a number of gluons radiated 
in the course of the gluon evolution. As it has been noted in~[28], on
the average the gluon transverse momentum decreases from the hard interaction
block towards the proton. We assume that the gluon 
emitted in the last evolution step and having the four-momenta $k'$ 
compensates the whole transverse momentum of the gluon participating in the hard 
subprocess, i.e. ${\mathbf k'}_{T} \simeq - {\mathbf k}_{T}$. All the other emitted gluons are 
collected together in the proton remnant, which is assumed\footnote{Note that such 
assumption is also used in the KMR formalism.} to carry only a negligible 
transverse momentum compared to ${\mathbf k'}_{T}$. This gluon gives rise to a 
final hadron jet with $E_T^{\rm jet} = |{\mathbf k'}_{T}|$ in addition to the jet 
produced in the hard subprocess. From these three hadron jets we choose the two ones 
carrying the largest transverse energies, and then compute the beauty and 
associated dijet photoproduction rates.

The recent experimental data~[6, 7] on the beauty and associated dijet
production at HERA come from both H1 and ZEUS collaborations.
The ZEUS data~[6] refer to the kinematical region defined by
$0.2 < y < 0.8$, $Q^2 < 1$~GeV$^2$ and given for jets with
$p_T^{{\rm jet}_1} > 7$~GeV, $p_T^{{\rm jet}_2} > 6$~GeV and
$|\eta^{\rm jet}| < 2.5$. The measured cross sections have been presented
for muons coming from semileptonic $b$ decays in dijet events with
$p_T^\mu > 2.5$~GeV and $-1.6 < \eta^\mu < 2.3$.
The more recent H1 data~[7] refer to the same kinematical region
except another muon pseudo-rapidity requirement: $-0.55 < \eta^\mu < 1.1$.
To produce muons from $b$-quarks in our theoretical calculations,
we first convert $b$-quarks into $B$-hadrons using
the Peterson fragmentation function~[45] and then 
simulate their semileptonic decay according to the
standard electroweak theory\footnote{Of course, the muon transverse momenta spectra are 
sensitive to the fragmentation functions. However, this dependence is 
expected to be small as compared with the uncertainties coming from the unintegrated 
gluon densities in a proton and in a photon.}. Our default set 
of the fragmentation parameter is $\epsilon_b = 0.0035$.

So, the transverse momentum and pseudo-rapidity distributions
of the $b$-quark decay muon for different kinematical region are shown
in Figs.~3 --- 6 in comparison to the HERA data. 
One can see that calculated cross sections (using the J2003 and KMS 
unintegrated gluon densities) agree very well with the experimental data
except the low $p_T^{\mu}$ region ($p_T^{\mu} < 3$~GeV) in Fig.~4. Note, however, that the 
behaviour of measured cross sections in this region is very different from each other 
in the H1 and ZEUS data. The $p_T^\mu$
distribution measured~[7] by the H1 collaboration falls steeply
with increasing transverse momentum $p_T^\mu$. 
The similar situation is observed also for the cross section measured as
a function of the transverse momentum of leading jet $p_T^{\rm jet}$ (see Fig.~7).
Our calculations give a less steep behaviour and are lower than the H1 data in the 
lowest momentum bin by a factor of 2.5. But at higher 
transverse momenta $p_T^\mu$ better agreement is obtained. 
Note that in the case of $p_T^{\rm jet}$ distribution the discrepancy 
at low $p_T^{\rm jet}$ is smaller, is about 1.5 times only.
In contrast, a good description of the ZEUS data~[6] (both in normalization and shape) 
for all values of $p_T^\mu$ is observed (see Fig.~3).
Therefore there is some inconsistency between the data.

Also the ZEUS collaboration have presented the data on the
transverse momentum and pseudo-rapidity distributions of
the jets associated with the muon (so-called $\mu$-jet) or $B$-hadron ($b$-jet).
These jets reproduce the kinematic of the $b$ (or $\bar b$) quark in a good
approximation. The $\mu$-jet is defined as the jet containing
the $B$-hadron that decays into the muon. Similarly, the $b$-jet is defined
as the jet containing the $B$ (or $\bar B$) hadron.
In Figs.~8 --- 11 we show our predictions for the
transverse momentum and pseudo-rapidity distributions of 
the $\mu$-jet and $b$-jet in comparison to the ZEUS measurements~[6].
One can see that J2003 and KMS gluon densities give results which 
agree well with the data, although slightly overestimate the data
at low $p_T^{\mu-{\rm jet}}$ (see Fig.~8).

We would like to note that the KMS gluon provides a more hard transverse momentum 
distribution of the final muon (or jets) as compared with other 
unintegrated densities under consideration. Similar
effect we have observed in the case of the inclusive beauty photoproduction (in a 
previous section).
Another interesting observation is that the dotted curves
which obtained using the KMR unintegrated gluon
lie below the H1 and ZEUS data everywhere. 
This fact confirms the assumption which was made in~[41] that 
the KMR formalism results in some underestimation of the 
calculated cross sections.
Also it is interesting that the difference in normalization between the KMS and J2003 predictions 
is rather small, is about 20\% only. 
However, it is in the contrast with the $D^*$ and dijet associated 
photoproduction at HERA which has been investigated in our previous paper~[30], where
we have found a relative large enhancement of the cross sections
calculated using the KMS gluon density. The 
possible explanation of this fact is that the 
large $b$-quark or $J/\psi$ meson mass (which provide a hard scale) 
make predictions of the perturbation theory of QCD more applicable.

Next we concentrate on the very interesting subject of study which is connected with the
individual contributions from the direct and resolved photon mechanisms
to the cross section in the $k_T$-factorization approach.
As it was already mentioned above, the $x_\gamma^{\rm obs}$ variable
(which corresponds at leading order to the fraction of the exchanged photon
momentum in the hard scattering process) provides a tool to
investigate the relative importance of different contributions.
In LO approximation, direct photon events at parton level have $x_\gamma^{\rm obs} \sim 1$, 
while the resolved photon events populate the low values of $x_\gamma^{\rm obs}$.
The same situation is observed in a NLO calculations, because in the three
parton final state any of these partons are allowed to take any kinematically
accessible value. In the $k_T$-factorization formalism the hardest transverse 
momentum parton emission can be anywhere 
in the evolution chain, and does not need to be closest to the photon as required by the 
strong $\mu^2$ ordering in DGLAP. Thus, if the two hardest jets are produced by the 
$b\bar b$ pair, then $x_\gamma^{\rm obs}$ is close to unity, but if a 
gluon from the initial cascade and one of the final $b$-quarks form
the two hardest transverse momentum jets, then $x_\gamma^{\rm obs} < 1$.
This statement is clearly demonstrated in Fig.~12 where separately shown the contributions
from the photon-gluon (dashed curve) and gluon-gluon fusion (dash-dotted curve) subprocesses. 
The solid curve represents the sum of both these contributions.
We have used here the KMR unintegrated gluon density for illustration.
As it was expected, the gluon-gluon fusion events (with a gluon coming from the photon) 
are distributed over the whole $x_\gamma^{\rm obs}$ range. It is clear that 
these events play important role at small values of $x_\gamma^{\rm obs}$.
Next, in agreement with the expectation for direct photon processes, the peak 
at high values of the $x_\gamma^{\rm obs}$ is observed.
However, one can see that off-shell photon-gluon fusion results 
also in substantial tail at small values of $x_\gamma^{\rm obs}$. 
The existence of this plateau in the collinear approximation of QCD usually is 
attributted to the heavy quark excitation from resolved photon.
In the $k_T$-factorization approach such plateau indicates the fact that the gluon 
radiated from evolution cascade appears to be harder than $b$-quarks 
(produced in hard parton interaction) in a significant fraction of events~[28--30].

In Fig.~13 and~14 we confront the $x_\gamma^{\rm obs}$ distributions
calculated in different kinematical regions with the HERA data~[6, 7].
One can see that the J2003 and KMR unintegrated gluon densities give a reasonable description
of the data but tend to slightly underestimate them
at middle and low $x_\gamma^{\rm obs}$. In the case of KMS gluon this
discrepancy is more significant since the contribution from the 
gluon-gluon fusion subprocess are not taken into account here.
The Monte-Carlo generator \textsc{Cascade}~[27] (which
generates low $x_\gamma^{\rm obs}$ events via initial state
gluon radiation without using a gluon density in a photon) also
underestimate~[7] the cross sections at low $x_\gamma^{\rm obs}$.
Note that the shapes of $x_\gamma^{\rm obs}$ distributions predicted
by the J2003 and KMR densities differs from each other.
This fact is connected with different properties of 
corresponding unintegrated gluon distributions in a proton and in a photon.
In general, from Fig.~12 --- 14 we can conclude that the gluon-gluon fusion contribution is 
important in description of the experimental data and that the behaviour 
of calculated $x_\gamma^{\rm obs}$ distributions at low values of $x_\gamma^{\rm obs}$ 
is strongly depends on the unintegrated gluon densities used.
However, our calculations still reasonable agree with the H1 and ZEUS data within the 
theoretical and experimental uncertainties.

\begin{table}
\begin{center}
\begin{tabular}{|l|c|}
\hline
  Source & $\sigma(ep \to e b\bar b + X \to e jj \mu + X')$~[pb] \\
\hline
  H1 measurement~[7] & $38.4~\pm~3.4$~(stat.)~$\pm~5.4$~(sys.) \\
  NLO QCD (FMNR)~[44] & $23.8^{+7.4}_{-5.1}$ \\
  \textsc{Cascade} (J2003 set 2) & 22.6 \\
  \textsc{Pythia}~[46] & 20.9 \\
  J2003 set 1 & 28.37 \\
  J2003 set 2 & 27.33 \\
  J2003 set 3 & 29.25 \\
  KMR & 17.43 \\
  KMS & 33.87 \\
  KMS ($m_b = 4.5$~GeV, $\Lambda_{\rm QCD} = 250$~MeV) & 38.84 \\
\hline
\end{tabular}
\end{center}
\caption{The total cross section of the beauty and associated dijet photoproduction 
obtained in the kinematic range 
$-0.55 < \eta^{\mu} < 1.1$, $p_T^{\mu} > 2.5$~GeV,
$Q^2 < 1$~GeV$^2$, $0.2 < y < 0.8$, $p_T^{{\rm jet}_1} > 7$~GeV, 
$p_T^{{\rm jet}_2} > 6$~GeV and $|\eta^{\rm jet}| < 2.5$.}
\end{table}

Now we turn to the total cross section of $b$-quark and associated dijet
photoproduction. The ZEUS collaboration 
has presented~[6] the results for forward, barrel and rear muon-chambers regions
which defined by $-1.6 < \eta^\mu < - 0.9$, $p_T^\mu > 2.5$~GeV (rear),
$-0.9 < \eta^\mu < 1.3$, $p_T^\mu > 2.5$~GeV (barrel) and
$1.48 < \eta^\mu < 2.3$, $p_T^\mu > 2.5$~GeV, $p^\mu > 4$~GeV (forward).
In Fig.~15 we display the results of our calculations in comparison to the recent ZEUS data.
One can see that our predictions (supplemented with the J2003 and KMS
gluon densities) agree well with the measurements
in the barrel region but underestimate the data in rear and forward ones.
The main discrepancy is found in forward kinematical region where our predictions
are below the data by a factor of 1.5. In Table~2 we
compare the calculated cross section with the H1 data~[7] which has been obtained 
in another kinematical region (defined above). The predictions
of Monte-Carlo programs \textsc{Pythia}~[46], \textsc{Cascade}~[27]
as well as NLO QCD calculations (FMNR)~[44] are also shown for comparison.
Note that these results are in a good agreement
with each other but are about 1.5 standard deviations below~[7] the data.
Our predictions are somewhat higher but still lie below the data, too.
However, this discrepancy is not dramatic, because 
some reasonable variations in beauty mass $m_b$, energy scale $\mu^2$ or 
$\Lambda_{\rm QCD}$ parameter,
namely $4.5 < m_b < 5$~GeV, $\mu_0^2/2 < \mu^2 < 2 \mu_0^2$ (where 
$\mu_0$ is the transverse mass of produced $b$-quark) and 
$150 < \Lambda_{\rm QCD} < 250$~MeV, can completely eliminate
the visible disagreement. To be precise, we have repeated our calculations
using the KMS gluon density with the $m_b = 4.5$~GeV and $\Lambda_{\rm QCD} = 250$~MeV.
We obtained the value $\sigma = 38.84$~pb which is close to
the experimental data point $\sigma = 38.4 \pm 3.4 \pm 5.4$~pb.

Finally, we would like to note that in according to the 
analysis~[6, 7] which was done by the H1 and ZEUS collaborations, in order to obtain a 
realistic comparison of the data and NLO QCD calculations the corrections for hadronisation 
should be taken into account in the predictions. The correction factors are 
typically $0.8 - 1.1$ depending on a bin~[6, 7]. These factors are not accounted 
for in our analysis.

\section{Conclusions} \indent 

We have investigated the beauty production in electron-proton
collisions at HERA in the $k_T$-factorization QCD approach.
The different photoproduction rates (both inclusive and associated with 
hadronic jets) have been studied. We took into account both the direct and resolved photon
contribution. In numerical analysis we have
used the unintegrated gluon densities which are obtained from the full CCFM, 
from unified BFKL-DGLAP evolution 
equations (KMS) as well as from the Kimber-Martin-Ryskin prescription.
Our investigations were based on LO off-mass shell 
matrix elements for the photon-gluon and gluon-gluon fusion subprocesses.
Special attention has been drawn to the $x_\gamma^{\rm obs}$ variable since this 
quantity is sensitive to relative contributions to the cross section from the 
different production mechanisms. We demonstrate the importance
of gluon-gluon fusion subprocess in description of the experimental data
at low values of $x_\gamma^{\rm obs}$.

We have shown that the $k_T$-factorization approach supplemented with the CCFM or 
BFKL-DGLAP evolved unintegrated gluon distributions (the J2003 or KMS densities) 
reproduces well the numerous HERA data
on beauty production. At the same time we have obtained that the
Kimber-Martin-Ryskin formalism results in some underestimation of the
cross sections. This shows the importance of a detail
understanding of the non-collinear parton evolution process.

\section{Acknowledgements} \indent 

The authors are very grateful to S.P.~Baranov for encouraging 
interest and very helpful discussions, L.K.~Gladilin for 
reading of the manuscript and very useful remarks.
This research was supported in part by the 
FASI of Russian Federation (grant NS-1685.2003.2).

\newpage

\begin{figure}
\epsfig{figure=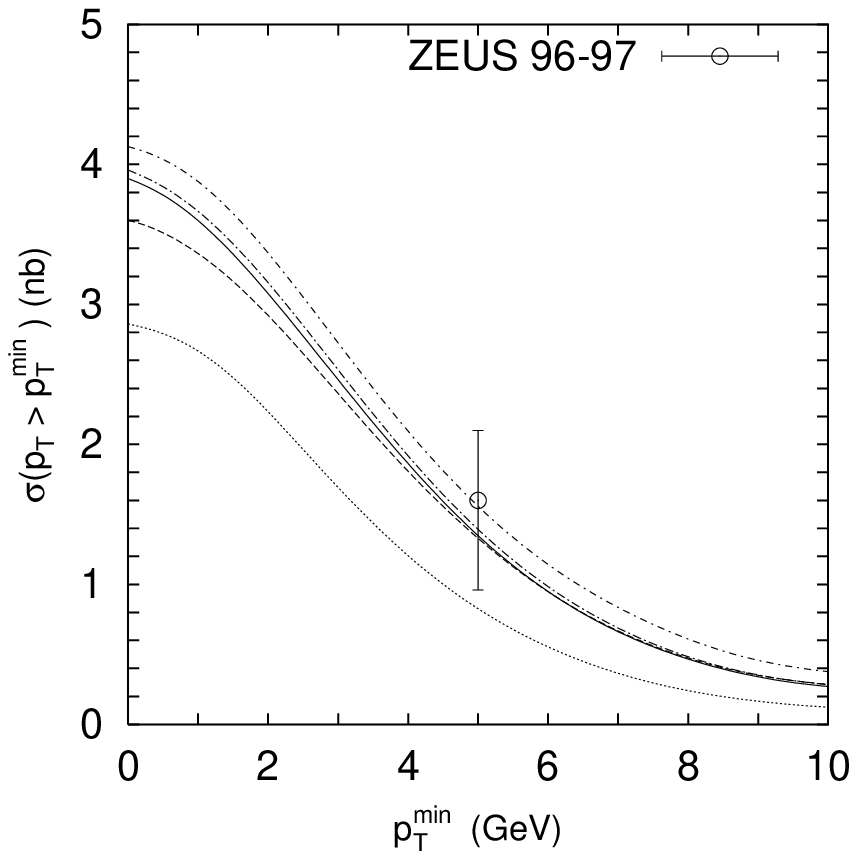, width = 22cm}
\caption{The inclusive beauty cross section as a function of $p_T^{\rm min}$ 
calculated at $|\eta^b| < 2$, $Q^2 < 1$~GeV$^2$ and $0.2 < y < 0.8$.
The solid, dashed, dash-dotted, dotted and short dash-dotted curves 
correspond to the J2003~set~1 --- 3, KMR and KMS unintegrated gluon 
distributions, respectively. The experimental data are from ZEUS~[5].}
\label{fig1}
\end{figure}

\newpage

\begin{figure}
\epsfig{figure=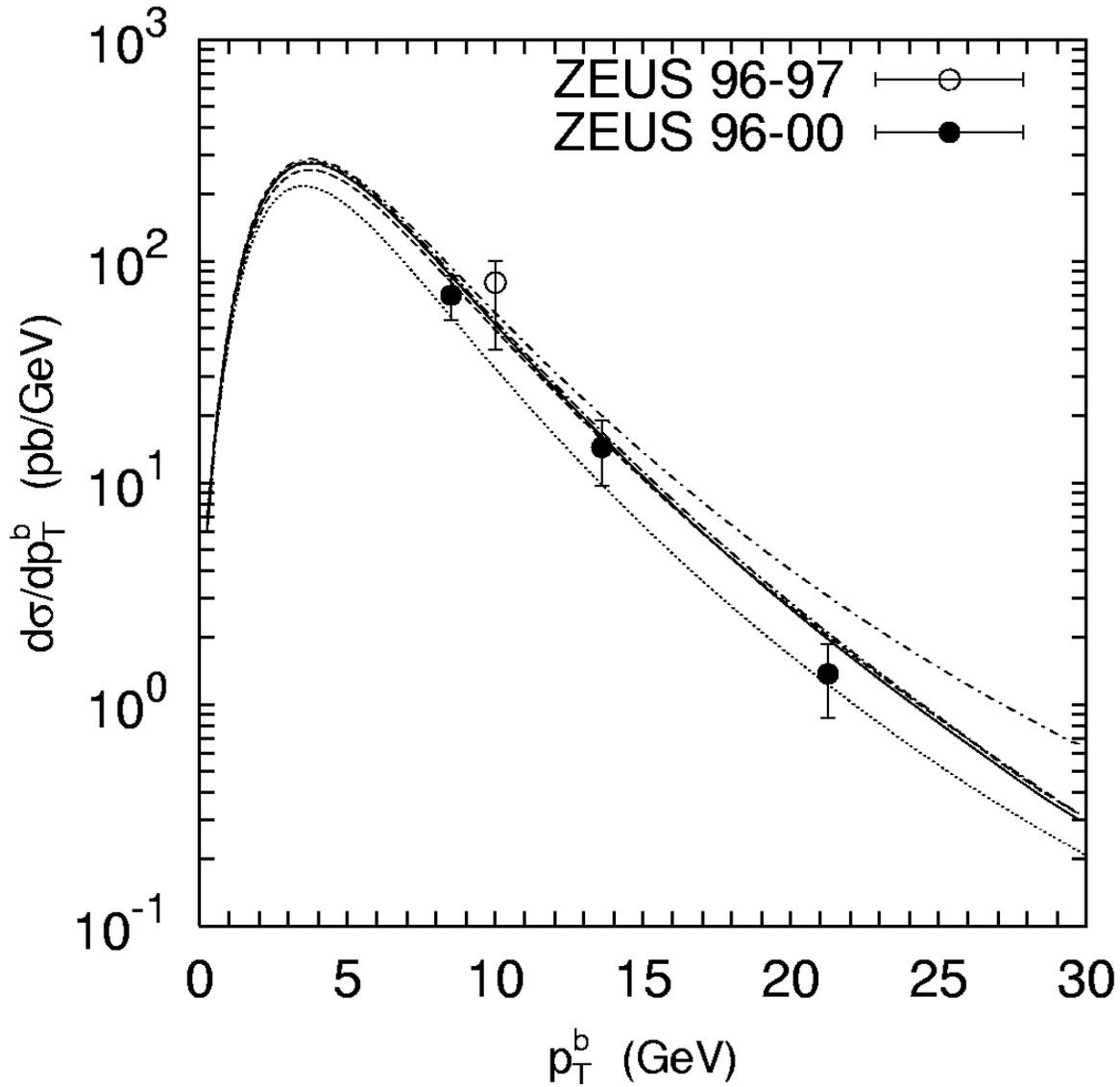, width = 22cm}
\caption{The differential cross section $d\sigma/d p_T^b$ for the inclusive 
beauty production calculated at $|\eta^b| < 2$, $Q^2 < 1$~GeV$^2$ and $0.2 < y < 0.8$.
All curves are the same as in Fig.~1. The experimental data are from ZEUS~[5, 6].}
\label{fig2}
\end{figure}

\newpage

\begin{figure}
\epsfig{figure=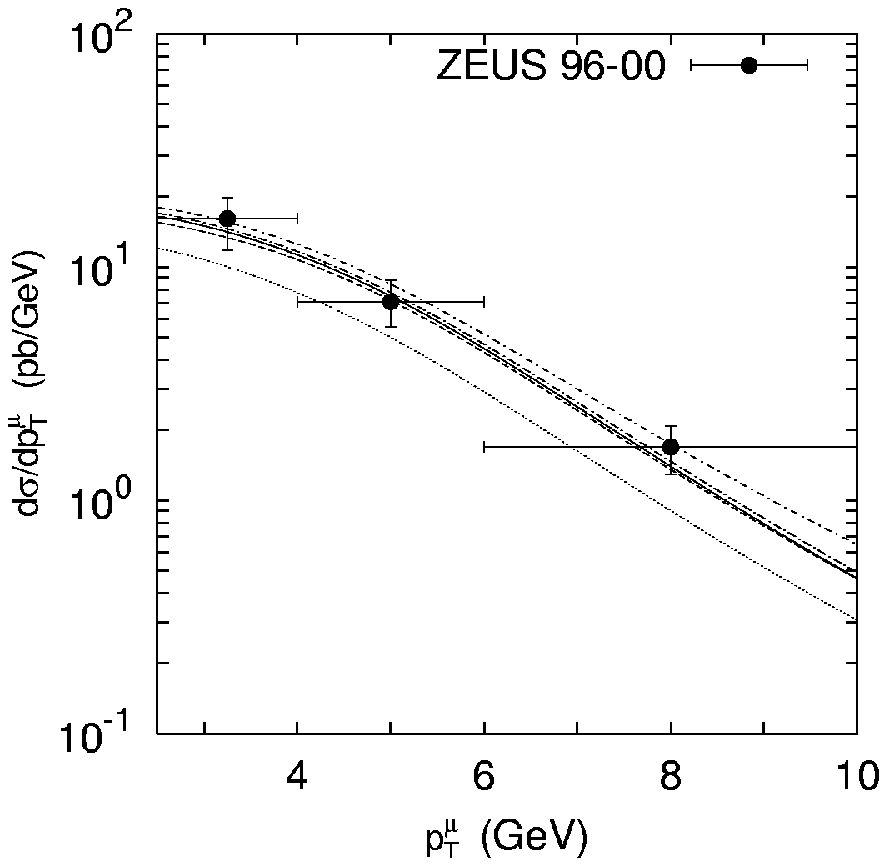, width = 22cm}
\caption{The differential cross section $d\sigma/d p_T^{\mu}$ for dijets with an 
associated muon coming from $b$ decays in the kinematic range $-1.6 < \eta^{\mu} < 2.3$, 
$Q^2 < 1$~GeV$^2$, $0.2 < y < 0.8$, $p_T^{{\rm jet}_1} > 7$~GeV, 
$p_T^{{\rm jet}_2} > 6$~GeV and $|\eta^{\rm jet}| < 2.5$.
All curves are the same as in Fig.~1. The experimental data are from ZEUS~[6].}
\label{fig3}
\end{figure}

\newpage

\begin{figure}
\epsfig{figure=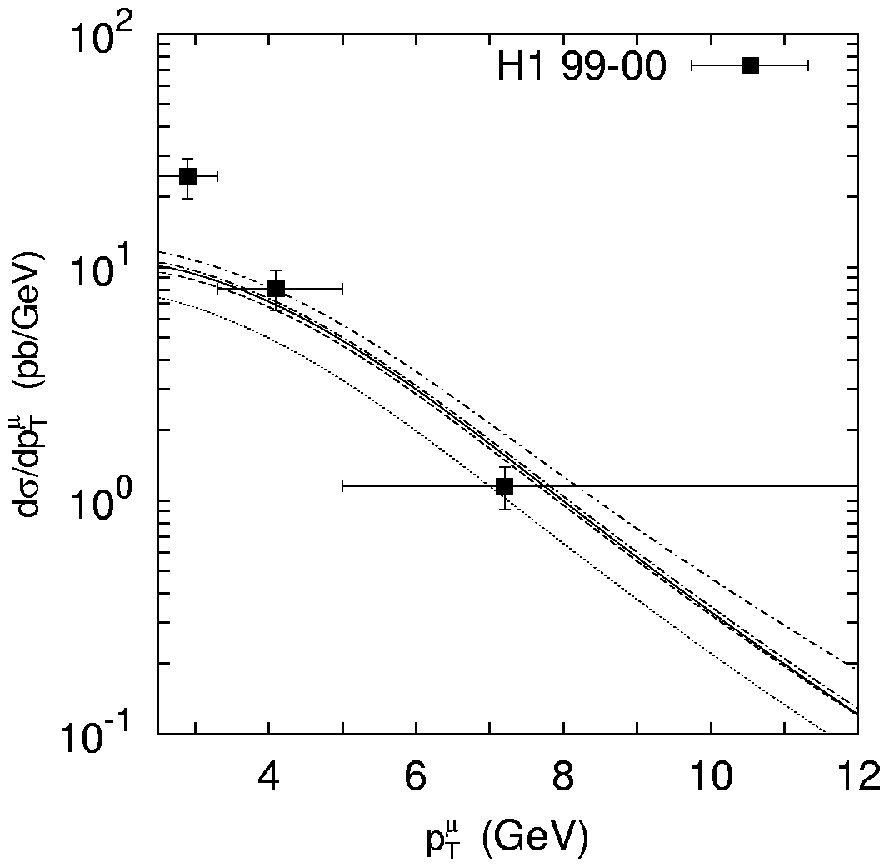, width = 22cm}
\caption{The differential cross section $d\sigma/d p_T^{\mu}$ for dijets with an 
associated muon coming from $b$ decays in the kinematic range $-0.55 < \eta^{\mu} < 1.1$, 
$Q^2 < 1$~GeV$^2$, $0.2 < y < 0.8$, $p_T^{{\rm jet}_1} > 7$~GeV, 
$p_T^{{\rm jet}_2} > 6$~GeV and $|\eta^{\rm jet}| < 2.5$.
All curves are the same as in Fig.~1. The experimental data are from H1~[7].}
\label{fig4}
\end{figure}

\newpage

\begin{figure}
\epsfig{figure=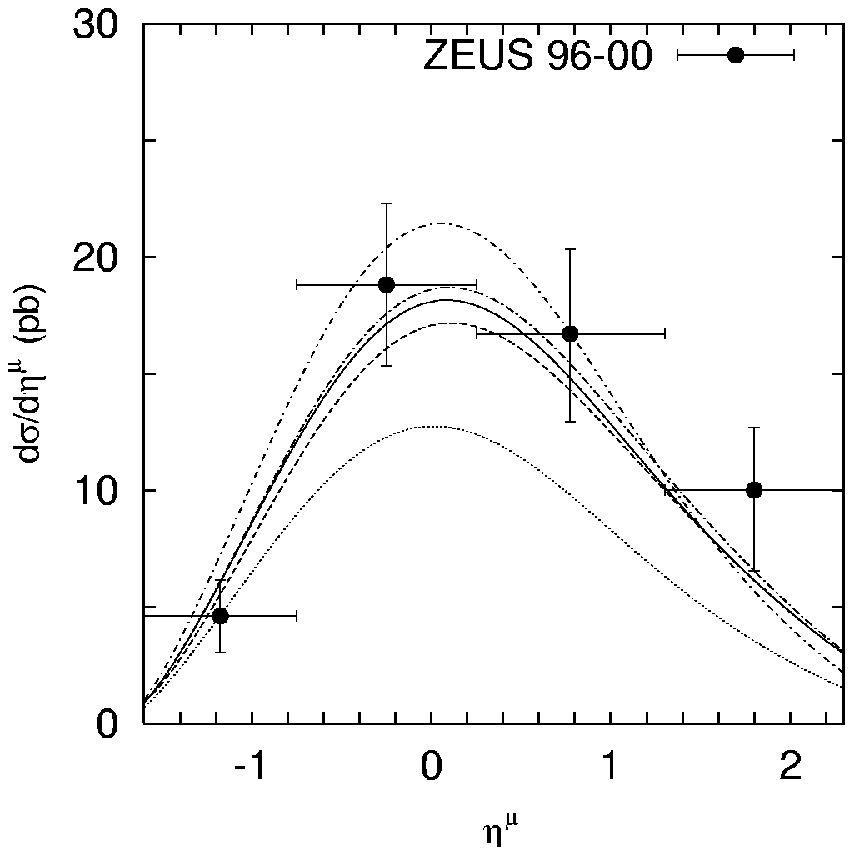, width = 22cm}
\caption{The differential cross section $d\sigma/d \eta^{\mu}$ for dijets with an 
associated muon coming from $b$ decays in the kinematic range $p_T^{\mu} > 2.5$~GeV, 
$Q^2 < 1$~GeV$^2$, $0.2 < y < 0.8$, $p_T^{{\rm jet}_1} > 7$~GeV, 
$p_T^{{\rm jet}_2} > 6$~GeV and $|\eta^{\rm jet}| < 2.5$.
All curves are the same as in Fig.~1. The experimental data are from ZEUS~[6].}
\label{fig5}
\end{figure}

\newpage

\begin{figure}
\epsfig{figure=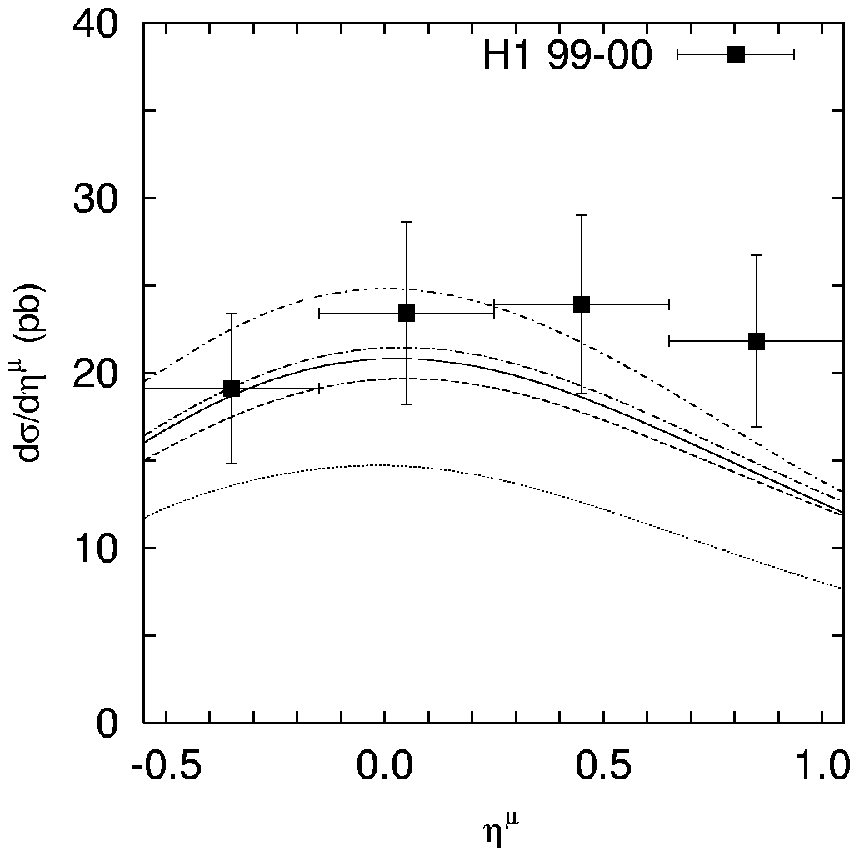, width = 22cm}
\caption{The differential cross section $d\sigma/d \eta^{\mu}$ for dijets with an 
associated muon coming from $b$ decays in the kinematic range $p_T^{\mu} > 2.5$~GeV, 
$Q^2 < 1$~GeV$^2$, $0.2 < y < 0.8$, $p_T^{{\rm jet}_1} > 7$~GeV, 
$p_T^{{\rm jet}_2} > 6$~GeV and $|\eta^{\rm jet}| < 2.5$.
All curves are the same as in Fig.~1. The experimental data are from H1~[7].}
\label{fig6}
\end{figure}

\newpage

\begin{figure}
\epsfig{figure=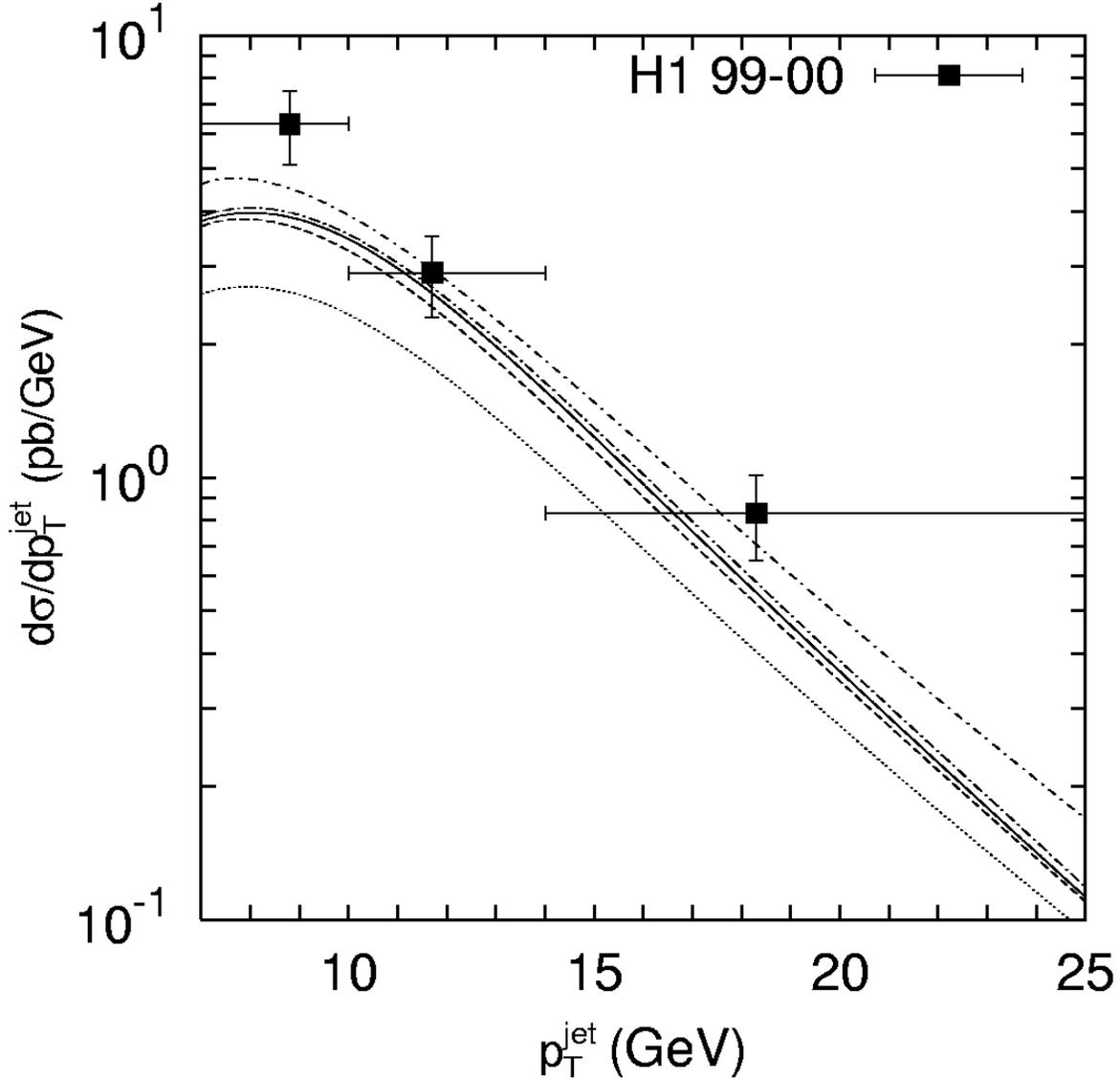, width = 22cm}
\caption{The leading jet transverse momentum distribution $d\sigma/d p_T^{\rm jet}$ for dijets with an 
associated muon coming from $b$ decays in the kinematic range $-0.55 < \eta^{\mu} < 1.1$, 
$Q^2 < 1$~GeV$^2$, $0.2 < y < 0.8$, $p_T^{{\rm jet}_1} > 7$~GeV, 
$p_T^{{\rm jet}_2} > 6$~GeV and $|\eta^{\rm jet}| < 2.5$.
All curves here are the same as in Fig.~1. The experimental data are from H1~[7].}
\label{fig7}
\end{figure}

\newpage

\begin{figure}
\epsfig{figure=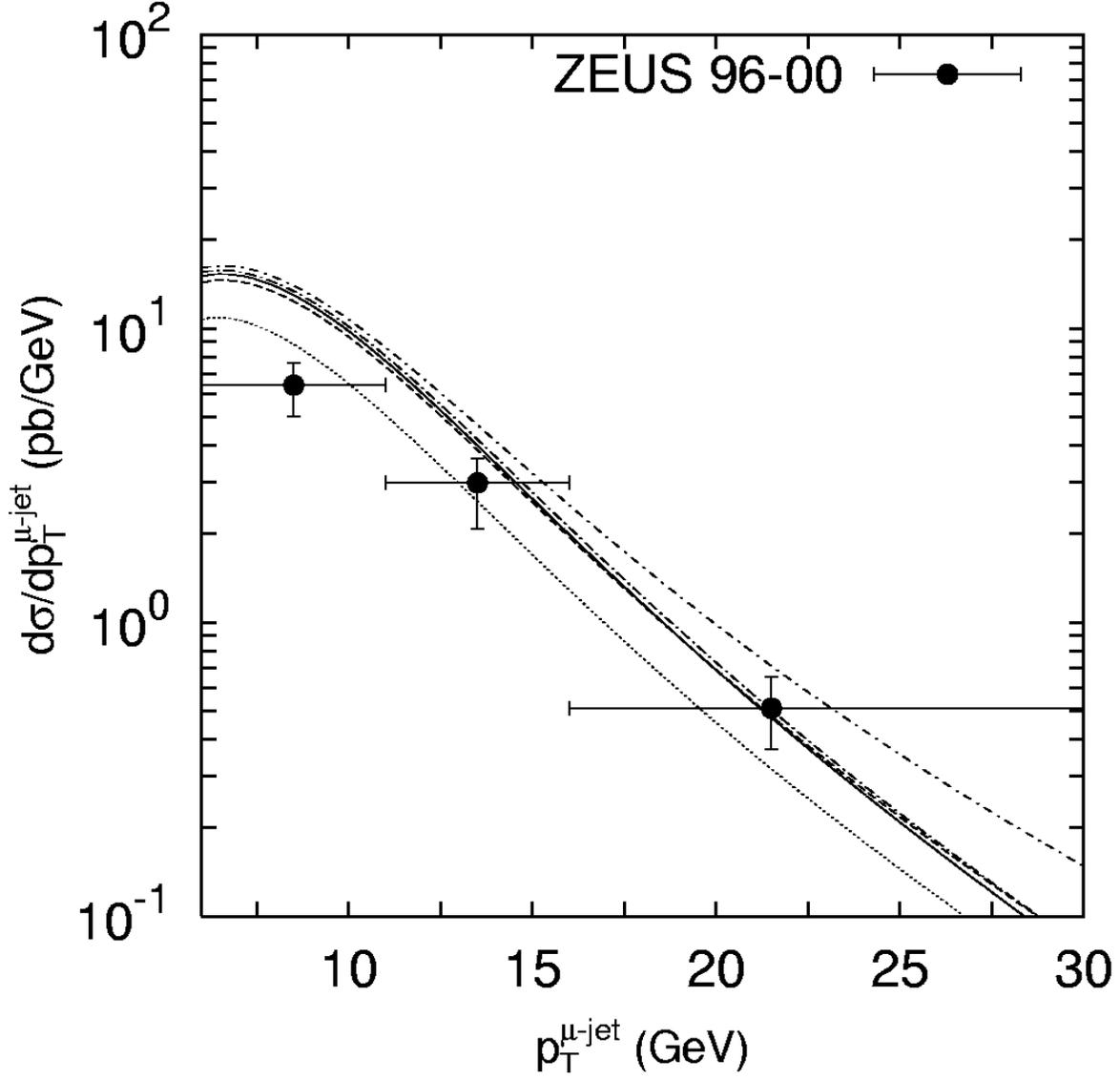, width = 22cm}
\caption{The transverse momentum distribution of the jet 
associated to the muon coming from $b$ decays in the kinematic range $-0.55 < \eta^{\mu} < 1.1$,
$p_T^{\mu} > 2.5$~GeV, $Q^2 < 1$~GeV$^2$, $0.2 < y < 0.8$ and $|\eta^{\mu-\rm jet}| < 2.5$.
All curves are the same as in Fig.~1. The experimental data are from ZEUS~[6].}
\label{fig8}
\end{figure}

\newpage

\begin{figure}
\epsfig{figure=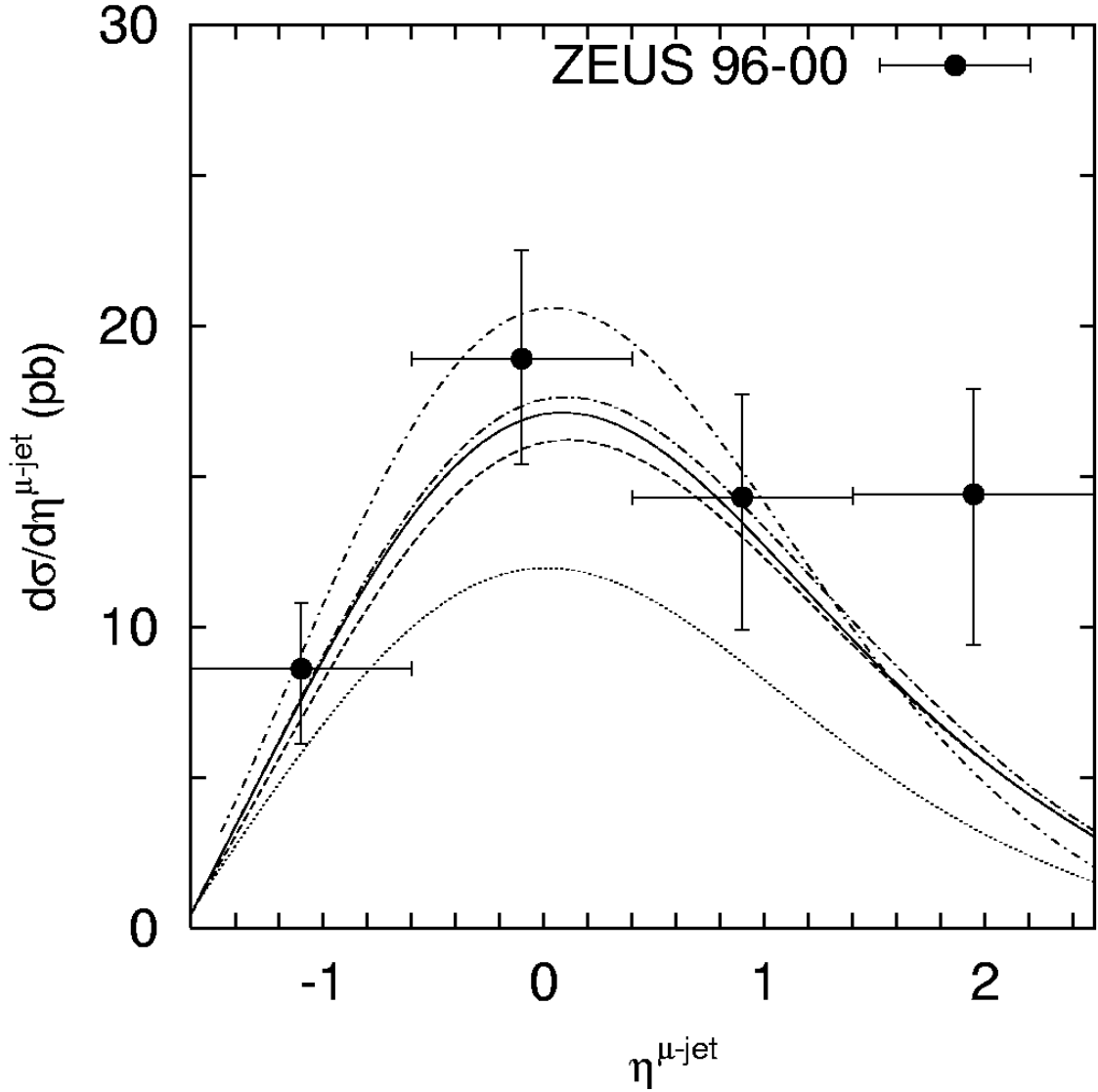, width = 22cm}
\caption{The pseudo-rapidity distribution of the jet 
associated to the muon coming from $b$ decays in the kinematic range $-0.55 < \eta^{\mu} < 1.1$,
$p_T^{\mu} > 2.5$~GeV, $Q^2 < 1$~GeV$^2$, $0.2 < y < 0.8$ and $p_T^{\mu-{\rm jet}} > 6$~GeV. 
All curves are the same as in Fig.~1. The experimental data are from ZEUS~[6].}
\label{fig9}
\end{figure}

\newpage

\begin{figure}
\epsfig{figure=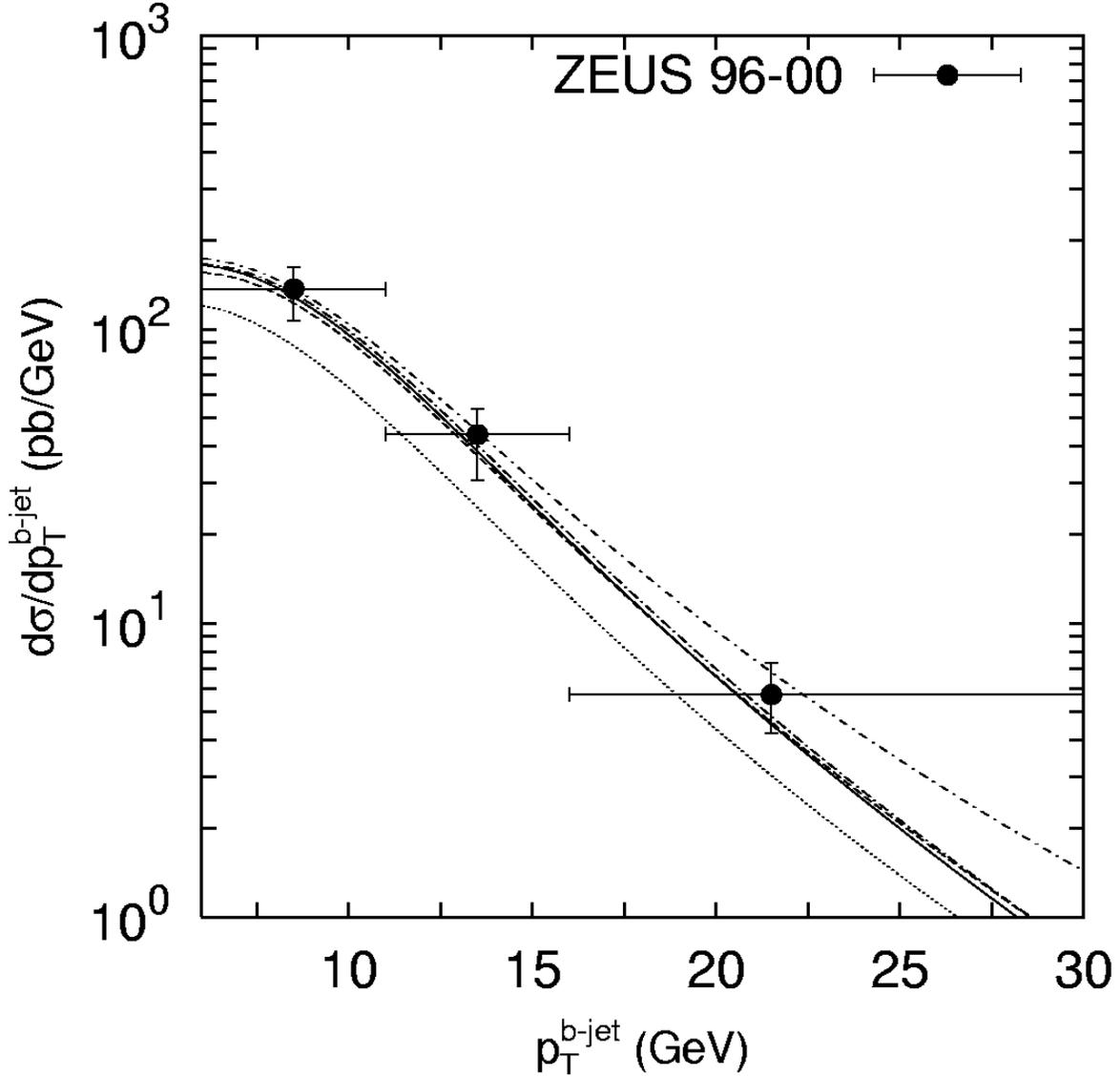, width = 22cm}
\caption{The transverse momentum distribution of the jet 
containing a $B$-hadron in the kinematic range $-0.55 < \eta^{\mu} < 1.1$,
$p_T^{\mu} > 2.5$~GeV, $Q^2 < 1$~GeV$^2$, $0.2 < y < 0.8$ and $|\eta^{b-\rm jet}| < 2.5$.
All curves are the same as in Fig.~1. The experimental data are from ZEUS~[6].}
\label{fig10}
\end{figure}

\newpage

\begin{figure}
\epsfig{figure=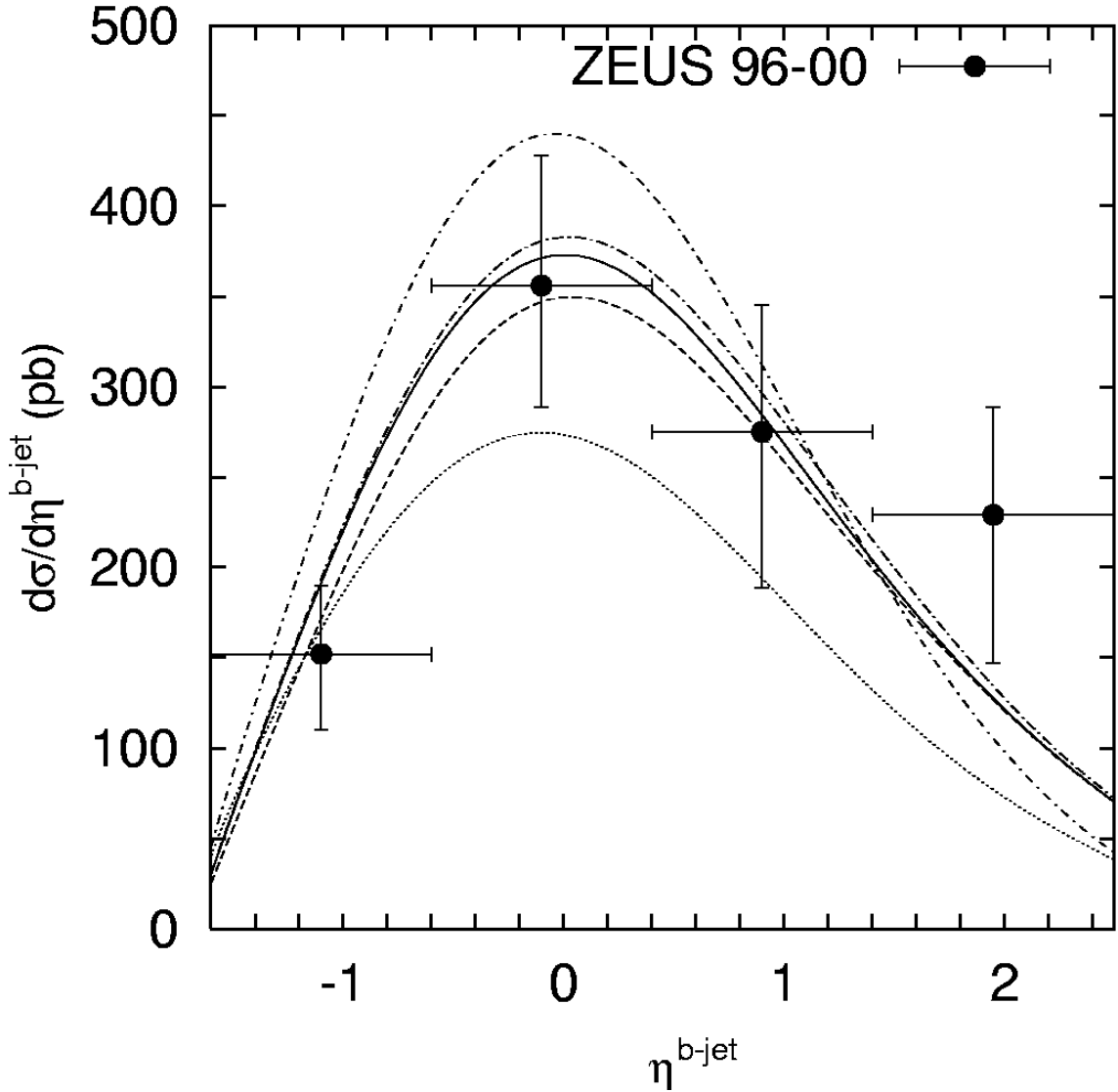, width = 22cm}
\caption{The pseudo-rapidity distribution of the jet 
containing a $B$-hadron in the kinematic range $-0.55 < \eta^{\mu} < 1.1$,
$p_T^{\mu} > 2.5$~GeV, $Q^2 < 1$~GeV$^2$, $0.2 < y < 0.8$ and $|\eta^{b-\rm jet}| < 2.5$.
All curves are the same as in Fig.~1. The experimental data are from ZEUS~[6].}
\label{fig11}
\end{figure}

\newpage

\begin{figure}
\epsfig{figure=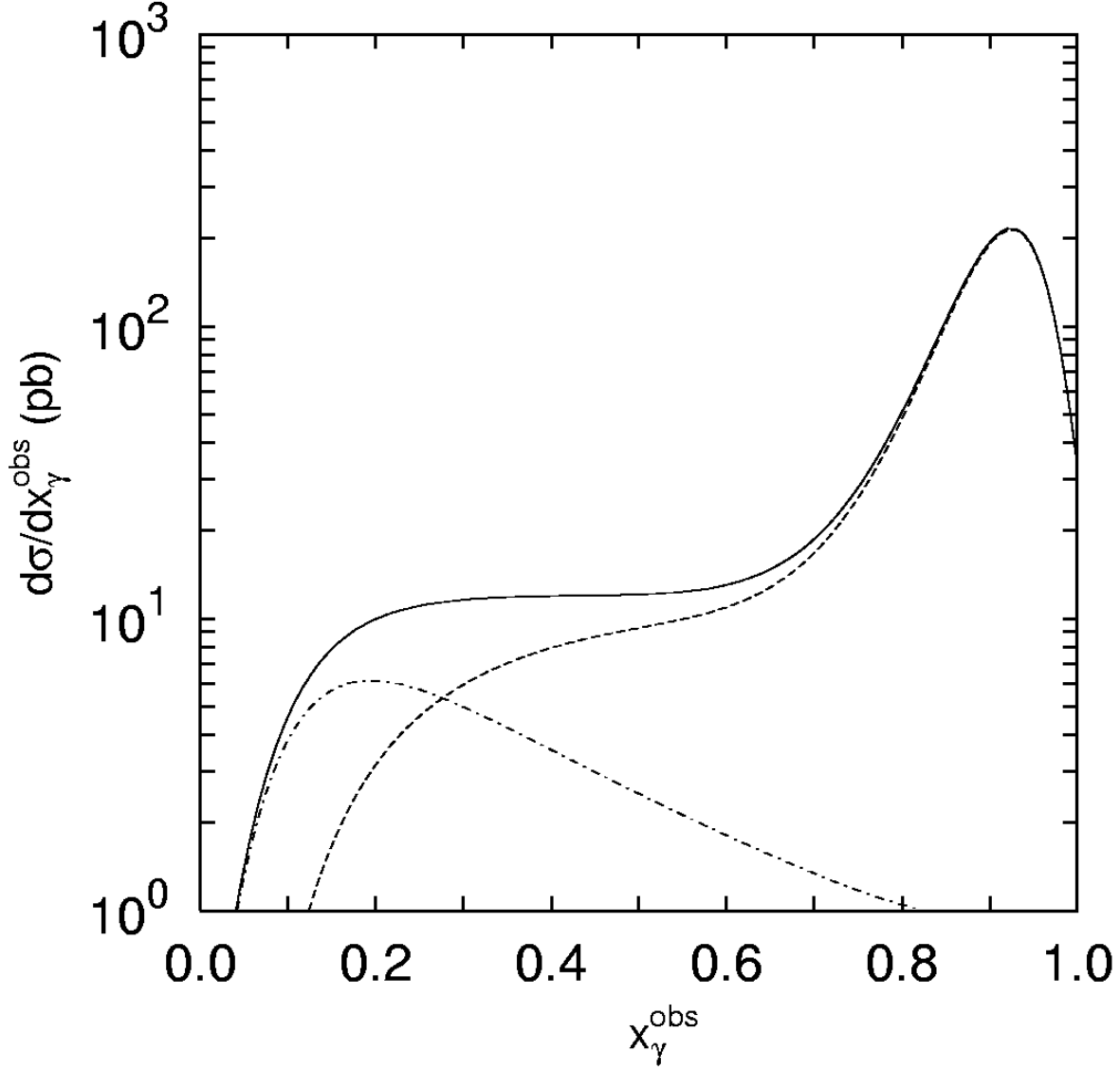, width = 22cm}
\caption{The differential cross section $d\sigma/d x_\gamma^{\rm obs}$ for dijets with an 
associated muon coming from $b$ decays in the kinematic range $-1.6 < \eta^{\mu} < 2.3$, 
$p_T^{\mu} > 2.5$~GeV, $Q^2 < 1$~GeV$^2$, $0.2 < y < 0.8$, $p_T^{{\rm jet}_1} > 7$~GeV, 
$p_T^{{\rm jet}_2} > 6$~GeV and $|\eta^{\rm jet}| < 2.5$.
Separately shown the contributions from the photon-gluon (dashed curve) and gluon-gluon 
fusion (dash-dotted curve). Solid curve represents the sum of both these contributions.
The KMR unintegrated gluon densities in a proton and in a photon has been used.}
\label{fig12}
\end{figure}

\newpage

\begin{figure}
\epsfig{figure=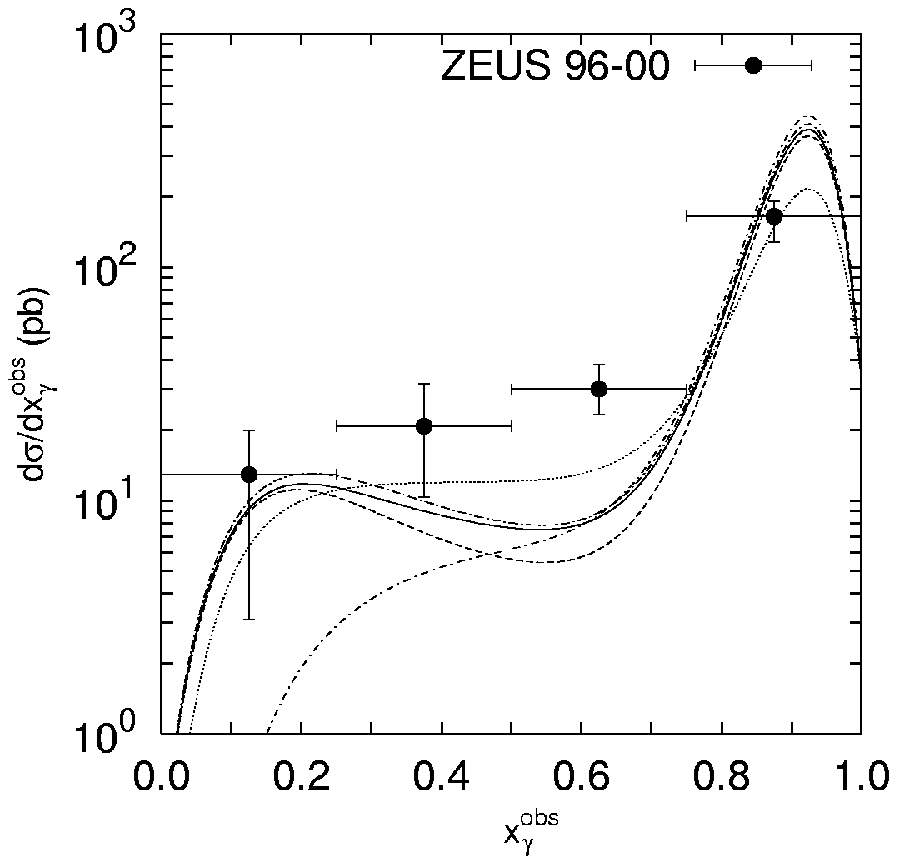, width = 22cm}
\caption{The differential cross section $d\sigma/d x_\gamma^{\rm obs}$ for dijets with an 
associated muon coming from $b$ decays in the kinematic range $-1.6 < \eta^{\mu} < 2.3$, 
$p_T^{\mu} > 2.5$~GeV, $Q^2 < 1$~GeV$^2$, $0.2 < y < 0.8$, $p_T^{{\rm jet}_1} > 7$~GeV, 
$p_T^{{\rm jet}_2} > 6$~GeV and $|\eta^{\rm jet}| < 2.5$.
All curves are the same as in Fig.~1. The experimental data are from ZEUS~[6].}
\label{fig13}
\end{figure}

\newpage

\begin{figure}
\epsfig{figure=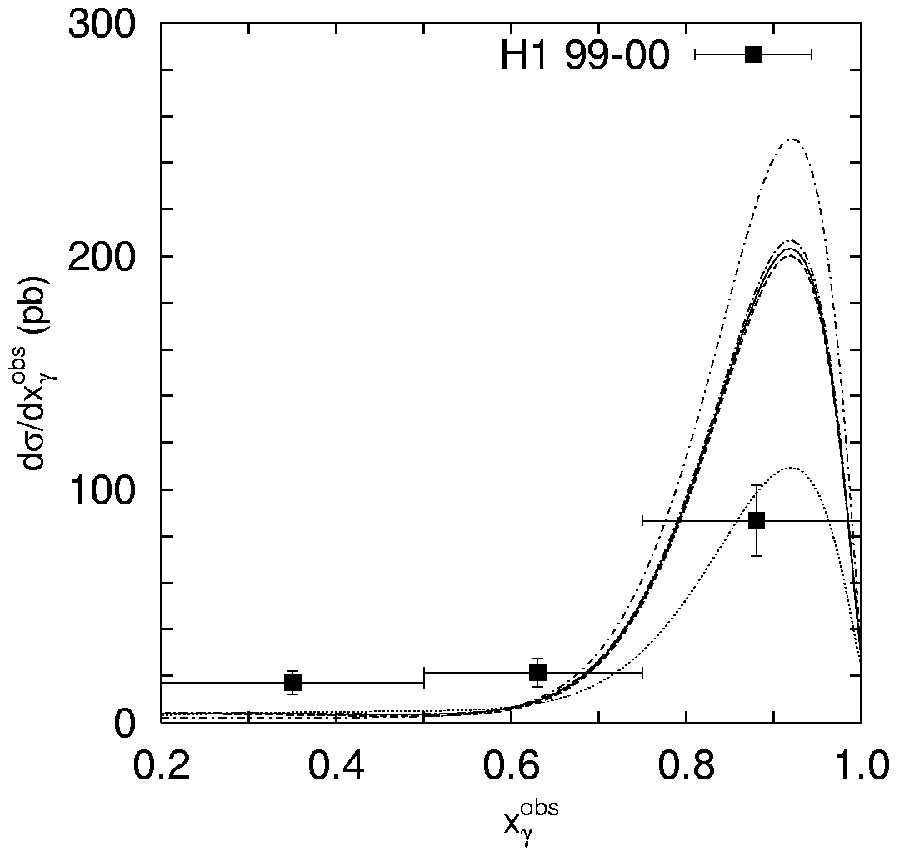, width = 22cm}
\caption{The differential cross section $d\sigma/d x_\gamma^{\rm obs}$ for dijets with an 
associated muon coming from $b$ decays in the kinematic range $-0.55 < \eta^{\mu} < 1.1$,
$p_T^{\mu} > 2.5$~GeV, $Q^2 < 1$~GeV$^2$, $0.2 < y < 0.8$, $p_T^{{\rm jet}_1} > 7$~GeV, 
$p_T^{{\rm jet}_2} > 6$~GeV and $|\eta^{\rm jet}| < 2.5$.
All curves are the same as in Fig.~1. The experimental data are from H1~[7].}
\label{fig14}
\end{figure}

\newpage

\begin{figure}
\epsfig{figure=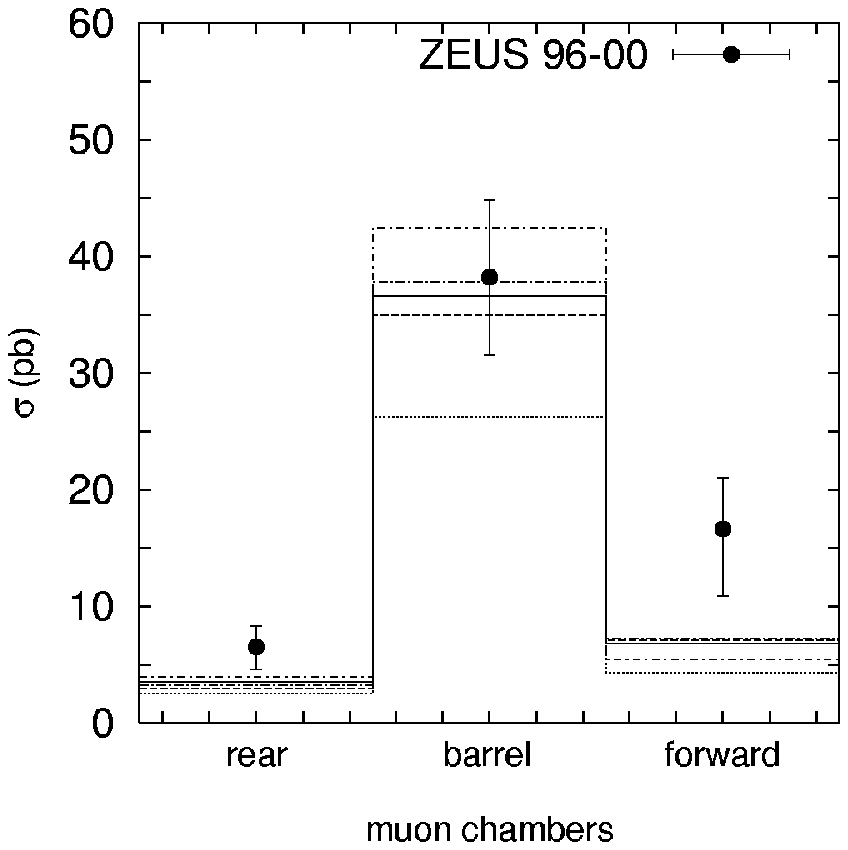, width = 22cm}
\caption{The cross section for muon coming from $b$ decays in dijet events 
calculated in the rear, barrel and forward kinematical regions (see text).
The cuts are applied: $Q^2 < 1$~GeV$^2$, $0.2 < y < 0.8$, $p_T^{{\rm jet}_1} > 7$~GeV, 
$p_T^{{\rm jet}_2} > 6$~GeV and $|\eta^{\rm jet}| < 2.5$.
All curves are the same as in Fig.~1. The experimental data are from ZEUS~[6].}
\label{fig15}
\end{figure}


\begin{thebibliography}{46}

\bibitem{1} C.~Adloff {\sl et al.} (H1 Collaboration), Phys. Lett. B {\bf 467}, 156 (1999); Erratum: {\it ibid} B {\bf 518}, 331 (2001).
\bibitem{2} F.~Abe {\sl et al.} (CDF Collaboration), Phys. Rev. D {\bf 55}, 2546 (1997);\\
  D.~Acosta {\sl et al.} (CDF Collaboration), Phys. Rev. D {\bf 65}, 052002 (2002);\\
  S.~Abachi {\sl et al.} (D0 Collaboration), Phys. Lett. B {\bf 487}, 264 (2000).
\bibitem{3} M.~Acciari {\sl et al.} (L3 Collaboration), Phys. Lett. B {\bf 503}, 10 (2001);\\
  P.~Achard {\sl et al.} (L3 Collaboration), Phys. Lett. B {\bf 619}, 71 (2005);\\
  G.~Abbiendi {\sl et al.} (OPAL Collaboration), Eur. Phys. J. C {\bf 16}, 579 (2000).
\bibitem{4} M.~Cacciari and P.~Nason, Phys. Rev. Lett. {\bf 89}, 122003 (2002);\\
  M.~Cacciari, S.~Frixione, M.L.~Mangano, P.~Nason, and G.~Ridolfi, JHEP {\bf 0407}, 033 (2004).
\bibitem{5} J.~Breitweg {\sl et al.} (ZEUS Collaboration), Eur. Phys. J. C {\bf 18}, 625 (2001).
\bibitem{6} S.~Chekanov {\sl et al.} (ZEUS Collaboration), Phys. Rev. D {\bf 70}, 012008 (2004).
\bibitem{7} A.~Aktas {\sl et al.} (H1 Collaboration), Eur. Phys. J. C {\bf 41}, 453 (2005).
\bibitem{8} S.~Catani, M.~Ciafoloni and F.~Hautmann, Nucl. Phys. B {\bf 366}, 135 (1991).
\bibitem{9} J.C.~Collins and R.K.~Ellis, Nucl. Phys. B {\bf 360}, 3 (1991).
\bibitem{10} L.V.~Gribov, E.M.~Levin, and M.G.~Ryskin, Phys. Rep. {\bf 100}, 1 (1983).
\bibitem{11} E.M.~Levin, M.G.~Ryskin, Yu.M.~Shabelsky and A.G.~Shuvaev, Sov. J. Nucl. Phys. {\bf 53}, 657 (1991).
\bibitem{12} E.A.~Kuraev, L.N.~Lipatov, and V.S.~Fadin, Sov. Phys. JETP {\bf 44}, 443 (1976);\\
  E.A.~Kuraev, L.N.~Lipatov, and V.S.~Fadin, Sov. Phys. JETP {\bf 45}, 199 (1977);\\
  I.I.~Balitsky and L.N.~Lipatov, Sov. J. Nucl. Phys. {\bf 28}, 822 (1978).
\bibitem{13} M.~Ciafaloni, Nucl. Phys. B {\bf 296}, 49 (1988);\\
  S.~Catani, F.~Fiorani, and G.~Marchesini, Phys. Lett. B {\bf 234}, 339 (1990);\\
  S.~Catani, F.~Fiorani, and G.~Marchesini, Nucl. Phys. B {\bf 336}, 18 (1990);\\
  G.~Marchesini, Nucl. Phys. B {\bf 445}, 49 (1995).
\bibitem{14} V.N.~Gribov and L.N.~Lipatov, Yad. Fiz. {\bf 15}, 781 (1972);\\
  L.N.~Lipatov, Sov. J. Nucl. Phys. {\bf 20}, 94 (1975);\\
  G.~Altarelly and G.~Parizi, Nucl. Phys. B {\bf 126}, 298 (1977);\\
  Y.L.~Dokshitzer, Sov. Phys. JETP {\bf 46}, 641 (1977).
\bibitem{15} M.G.~Ryskin and Yu.M.~Shabelsky, Z. Phys. C {\bf 61}, 517 (1994);\\
  M.G.~Ryskin, Yu.M.~Shabelsky and A.G.~Shuvaev, Z. Phys. C {\bf 69}, 269 (1996).
\bibitem{16} S.P.~Baranov and M.~Smizanska, Phys. Rev. D {\bf 62}, 014012 (2000).
\bibitem{17} Ph.~H\"agler, R.~Kirschner, A.~Sch\"afer, L.~Szymanowski and O.V.~Teryaev, Phys. Rev. D {\bf 62}, 071502 (2000).
\bibitem{18} H.~Jung, Phys. Rev. D {\bf 65}, 034015 (2002).
\bibitem{19} A.V.~Lipatov, N.P.~Zotov, and V.A.~Saleev, Yad. Fiz. {\bf 66}, 786 (2003);\\
  S.P.~Baranov, N.P.~Zotov and A.V.~Lipatov, Phys. Atom. Nucl. {\bf 67}, 834 (2004).
\bibitem{20} A.V.~Lipatov, L.~L\"onnblad, and N.P.~Zotov, JHEP {\bf 01}, 010 (2004).
\bibitem{21} H.~Jung, Mod. Phys. Lett. A {\bf 19}, 1 (2004).
\bibitem{22} L.~Motyka and N.~Timneanu, Eur. Phys. J. {\bf C27}, 73 (2003).
\bibitem{23} M.~Hansson, H.~Jung, and L.~J\"onsson, hep-ph/0402019.
\bibitem{24} A.V.~Lipatov and N.P.~Zotov, Eur. Phys. J. C {\bf 41}, 163 (2005).
\bibitem{25} A.V.~Lipatov, to be published in Yad. Fiz. (2006).
\bibitem{26} H.~Jung and G.~Salam, Eur. Phys. J. C {\bf 19}, 351 (2001).
\bibitem{27} H.~Jung, Comput. Phys. Comm. {\bf 143}, 100 (2002).
\bibitem{28} S.P.~Baranov and N.P.~Zotov, Phys. Lett. B {\bf 491}, 111 (2000).
\bibitem{29} S.P.~Baranov, H.~Jung, L.~J\"onsson, S.~Padhi, and N.P.~Zotov, Eur. Phys. J. C {\bf 24}, 425 (2002).
\bibitem{30} A.V.~Lipatov and N.P.~Zotov, DESY 05-252 [hep-ph/0512013].
\bibitem{31} J.~Kwiecinski, A.D.~Martin and A.M.~Stasto, Phys. Rev. D {\bf 56}, 3991 (1997).
\bibitem{32} M.A.~Kimber, A.D.~Martin and M.G.~Ryskin, Phys. Rev. D {\bf 63}, 114027 (2001);\\
  G.~Watt, A.D.~Martin and M.G.~Ryskin, Eur. Phys. J. C {\bf 31}, 73 (2003).
\bibitem{33} G.P.~Lepage, J. Comput. Phys. {\bf 27}, 192 (1978).
\bibitem{34} B.~Andersson {\sl et al.} (Small-$x$ Collaboration), Eur. Phys. J. C {\bf 25}, 77 (2002).
\bibitem{35} J.~Andersen {\sl et al.} (Small-$x$ Collaboration), Eur. Phys. J. C {\bf 35}, 77 (2004).
\bibitem{36} J.~Kwiecinski, A.D.~Martin and A.~Sutton, Phys. Rev. D {\bf 52}, 1445 (1995).
\bibitem{37} J.~Kwiecinski, A.D.~Martin and J. Outhwaite,  Eur. Phys. J. C {\bf 9}, 611 (2001).
\bibitem{38} A.V.~Lipatov and N.P.~Zotov, Eur. Phys. J. C {\bf 27}, 87 (2003).
\bibitem{39} N.P.~Zotov, I.I.~Katkov, and A.V.~Lipatov, to be published in Yad. Fiz (2006).
\bibitem{40} M.~Gl\"uck, E.~Reya and A.~Vogt, Phys. Rev. {\bf D46}, 1973 (1992);\\
  M.~Gl\"uck, E.~Reya and A.~Vogt, Z. Phys. {\bf C67}, 433 (1995). 
\bibitem{41} A.V.~Lipatov and N.P.~Zotov, Phys. Rev. D {\bf 72}, 054002 (2005).  
\bibitem{42} M.A.~Kimber, A.D.~Martin and M.G.~Ryskin, Eur. Phys. J. C {\bf 12}, 655 (2001).
\bibitem{43} A.V.~Lipatov and N.P.~Zotov, DESY 05-157 [hep-ph/0507243].
\bibitem{44} S.~Frixione, P.~Nason, and G.~Ridolfi, Nucl. Phys. B {\bf 454}, 3 (1995).
\bibitem{45} C.~Peterson, D.~Schlatter, I.~Schmitt, and P.~Zerwas, Phys. Rev. D {\bf 27}, 105 (1983).
\bibitem{46} T.~Sj\"ostrand {\sl et al.} Comput. Phys. Comm. {\bf 135}, 238 (2001).

\end{thebibliography}
\end{document}